\documentclass{aa}
\usepackage{graphics}
\usepackage{natbib}
\bibpunct{(}{)}{;}{a}{}{,}

\DeclareRobustCommand{\ion}[2]{%
\relax\ifmmode
\ifx\testbx\f@series
{\mathbf{#1\,\mathsc{#2}}}\else
{\mathrm{#1\,\mathsc{#2}}}\fi
\else\textup{#1\,{\mdseries\textsc{#2}}}%
\fi}

\def\apjl{ApJ}%
\def\apjs{ApJS}%
\def\ao{Appl.~Opt.}%
\def\aap{A\&A}%
\title{High resolution radio continuum survey of M33\\I. The radio maps }

\author{ F. S. Tabatabaei\thanks{Member of the International Max Planck Research School (IMPRS) for Radio and Infrared Astronomy at the Universities of Bonn and Cologne}, M. Krause, R. Beck }
\institute{Max-Planck Institut f\"ur Radioastronomie, Auf dem H\"ugel 69, 53121 Bonn, Germany   }
\offprints{F. Tabatabaei\\ tabataba@mpifr-bonn.mpg.de}

\begin{document}

\titlerunning{ High resolution radio continuum survey of M33}
\authorrunning{Tabatabaei et al.}
\abstract
{}
{We study the exponential scale length of total radio emission, the spectral index distribution, and the linear radio polarization in the Scd galaxy M33. }
{Observations  were carried out using the 3.6\,cm dual channel and the 6.2\,cm four channel receivers of the 100--m Effelsberg telescope along with the L-band VLA D--array at 20\,cm.  }
{High spatial resolution and sensitivity in both total and linearly polarized radio continuum emission from M33 were achieved. We found considerable extended emission, not only from the main arms I\,S and I\,N, but also from the weaker arms.   The large--scale magnetic field exhibits well--ordered spiral structure with almost the same orientation as that of the optical spiral arms, however, it does not show a clear structural correlation or anti--correlation with the optical arms. There is a north-south asymmetry in polarization that is frequency-dependent.  We found that the ring mean spectral index versus  radius  increases faster beyond $R$\,=\,4\,kpc. At each wavelength, the exponential scale length is larger inside than outside $R$\,=\,4\,kpc. }
{From the larger scales lengths at $R$\,$<$\,4\,kpc, we conclude  that star forming regions are mainly spread over the region $R$\,$<$\,4\,kpc without a dominant nuclear concentration. Furthermore, at $R$\,$<$\,4\,kpc, a spatial correlation between cosmic rays and star forming regions may exist. From the behaviour of the mean spectral indices obtained from  different pairs of the radio continuum data at 3.6, 6.2, and 20\,cm, we confirm that a decrease in the thermal fraction causes an increase in the spectral index.  The frequency-dependent asymmetry in the polarization hints to an asymmetry in Faraday depolarization.
\keywords{galaxies: individual: M33 -- radio continuum: galaxies -- galaxies: magnetic field -- galaxies: ISM }
}
\maketitle


\section{Introduction}

M33\,(NGC\,598), the nearest Scd galaxy at a distance of 840 kpc \citep[1$\arcsec \simeq$ 4\,pc, ][]{Freedman_etal_91}, has been extensively studied at radio wavelengths. \cite{Terzian} detected some diffuse emission at 318 and 606\,MHz using angular resolutions of 17$\arcmin$ and 10$\arcmin$, respectively. \cite{Israel_74} observed M33 at 1415\,MHz with the Synthesis telescope in Westerbork (WSRT). However, they did not find any extended spiral arm emission. This was first revealed by observations with the 100--m Effelsberg dish at 6.2\,cm \citep{vonKap-Herr} that was extensively studied by \cite{Berkhuijsen_83}. Radio polarization observations of M33 were first made by \cite{Beck_79} at 21.1\,cm and 11.1\,cm. \cite{Buczilowski_87} observed this galaxy at 17.4\,cm, 6.3\,cm, and 2.8\,cm. They detected polarization mostly in the northern half of M33 and proposed a bisymmetric magnetic field structure. The next interferometer observations \citep{Duric,viallefond_et_al_98,Gordons_99} were carried out in total intensity to study point sources like HII regions and supernova remnants (SNRs) within M33. 

Generally, single-dish observations should be carried out to study the extended radio emission.  
So far,  single--dish observations have been limited by sensitivity because M33 is relatively weak in radio emission, especially in polarized emission \citep[with an integrated flux density of $\sim$\,83\,mJy at 6.2\,cm, ][]{Buczilowski_etal_91} and because of M33's large extent in the sky ($\sim 35\arcmin \times 40\arcmin$).

This paper is the first of a series providing a detailed investigation of the radio continuum emission of M33. 
We describe new observations of the radio continuum emission at 3.6\,cm, 6.2\,cm, and 20\,cm with much improved detection limits of total and polarized intensity  and with higher spatial resolutions. These data allow a much more detailed study of polarization and  magnetic field within M33 than was possible before. We discuss the distribution and degree of linear polarization, integrated flux densities and distribution of the total spectral index  (obtained from total intensities of the radio continuum emission including both thermal and non-thermal emission). We also compare the radial,  exponential scale lengths of the radio emission at different wavelengths.

A comprehensive treatment of the separation of thermal and non-thermal emission, rotation measures, Faraday effects, depolarization, magnetic field strength and structure, radio--IR correlation, and cosmic ray distribution will be given in  further papers of this series. 

The observations and data reduction are described in Sect. 2. Results including distributions of total intensity and polarization, integrated flux densities and total spectral index distribution are discussed in Sect. 3. Discussion and concluding remarks are presented in Sect. 4.

\begin{table}
\begin{center}
\caption{Positional data on M33.}
\begin{tabular}{ l l } 
\hline
\hline
Nucleus position\,(J2000)$^{1}$    & RA\,=\,$1^{h}33^{m}51.0^{s}$      \\
    &  DEC\,=\,$30^{\circ}39\arcmin37.0\arcsec$\\
Position angle of major axis$^{2}$   &23$^{\circ}$ \\
Inclination$^{3}$    & 56$^{\circ}$ \\
Distance$^{4}$\,(1$\arcsec$=\,4\,kpc)   & 840\,kpc\\
\hline
\noalign {\medskip}
\multicolumn{2}{l}{$^{1}$ \cite{devaucouleurs_81}}\\
\multicolumn{2}{l}{$^{2}$ \cite{Deul}}\\
\multicolumn{2}{l}{$^{3}$ \cite{Regan_etal_94}}\\
\multicolumn{2}{l}{$^{4}$ \cite{Freedman_etal_91}}\\
\end{tabular}
\end{center}
\end{table}

\section{Observations and data reduction}

We performed both single-dish and interferometer observations. The single-dish observations at 3.6\,cm and 6.2\,cm were made with the 100-m Effelsberg telescope\footnote{The 100--m telescope at Effelsberg is operated by the Max-Planck-Institut f\"ur Radioastronomie (MPIfR) on behalf on the Max--Planck--Gesellschaft.}. The interferometer observations at 20\,cm  were carried out with the Very Large Array (VLA\footnote{The VLA is a facility of the National Radio Astronomy Observatory. The NRAO is operated by Associated Universities, Inc., under contract with the National Science Foundation.}). 

\subsection{Effelsberg single-dish observations }

We observed M33 at 3.6\,cm during several periods from August 2005 to March 2006 with the broadband system (bandwidth\,$\sim$\,1100\,MHz) centered at 8350\,MHz. The receiver is a dual-channel correlation radiometer with cooled HEMT amplifiers and a cooled polarization transducer located in the secondary focus of the 100-m telescope. The system is set up for receiving Left Hand Circular (LHC) and Right Hand Circular (RLC) polarization.  
M33 was scanned along RA and DEC with a scanning velocity of 50$\arcmin$/min on a grid of 30$\arcsec$.  The scanned area of 40$\arcmin$\,$\times$\,50$\arcmin$ was centered on the nucleus of the galaxy. The total observing time was about 100 hours. We obtained 56 coverages with a typical sensitivity per scan of 2\,mJy/beam area in total intensity and 0.4\,mJy/beam area in polarization. The Half Power Beam Width (HPBW) at 3.6\,cm is $83.6\arcsec$ which corresponds to a linear resolution of about 335\,pc.

The sources 3C\,48, 3C\,286, and 3C\,138 were used for pointing and focusing during the observations. Calibration of the flux density and polarization angle was achieved by  observing 3C\,286. We used the flux density scale given by \cite{Baars_etal_77}.
\begin{table}
\caption{Characteristics and performances of 3.6\,cm and 6.2\,cm receivers of the 100-m Effelsberg telescope.}
\begin{tabular}{ l l l } 
\hline
$\lambda$ (cm)& 3.6& 6.2  \\
\hline 
\hline
Center frequency (GHz)   & 8.350 & 4.850 \\
Band width (MHz)    & 1100  & 500 \\
HPBW ($\arcsec$)   & 83.6 & 146\\
$T_{sys}$ (K)    & 23-25 & 25-28 \\
$T_{b}/S_{\nu}$ (K/Jy)  & 2.62 & 2.4 \\
\hline
\end{tabular}
\end{table}

The data reduction was performed in the NOD2 data reduction system ~\citep{Haslam}. In order to remove the scanning effects due to ground radiation, weather condition and receiver instabilities, we applied the scanning removal program, $\it{Presse}$, of \cite{Sofue_etal_79}. The r.m.s. noise after combination of the coverages \citep{Emerson_etal_88} is $\sim$\,220\,$\mu$Jy/beam in Stokes I (total intensity) map and $\sim$\,70\,$\mu$Jy/beam in Stokes U and Q (polarization) maps. The final maps in U and Q were combined to produce maps in polarized intensity and in polarization angle, correcting for the positive bias in polarized intensity due to noise \citep{Killeen}.
Figs. 1 and 2  show the resultant maps of the total and polarized intensities, respectively, smoothed to an angular resolution of 120$\arcsec$. 
\begin{figure*}
\sidecaption
\resizebox{12cm}{!}{\includegraphics*{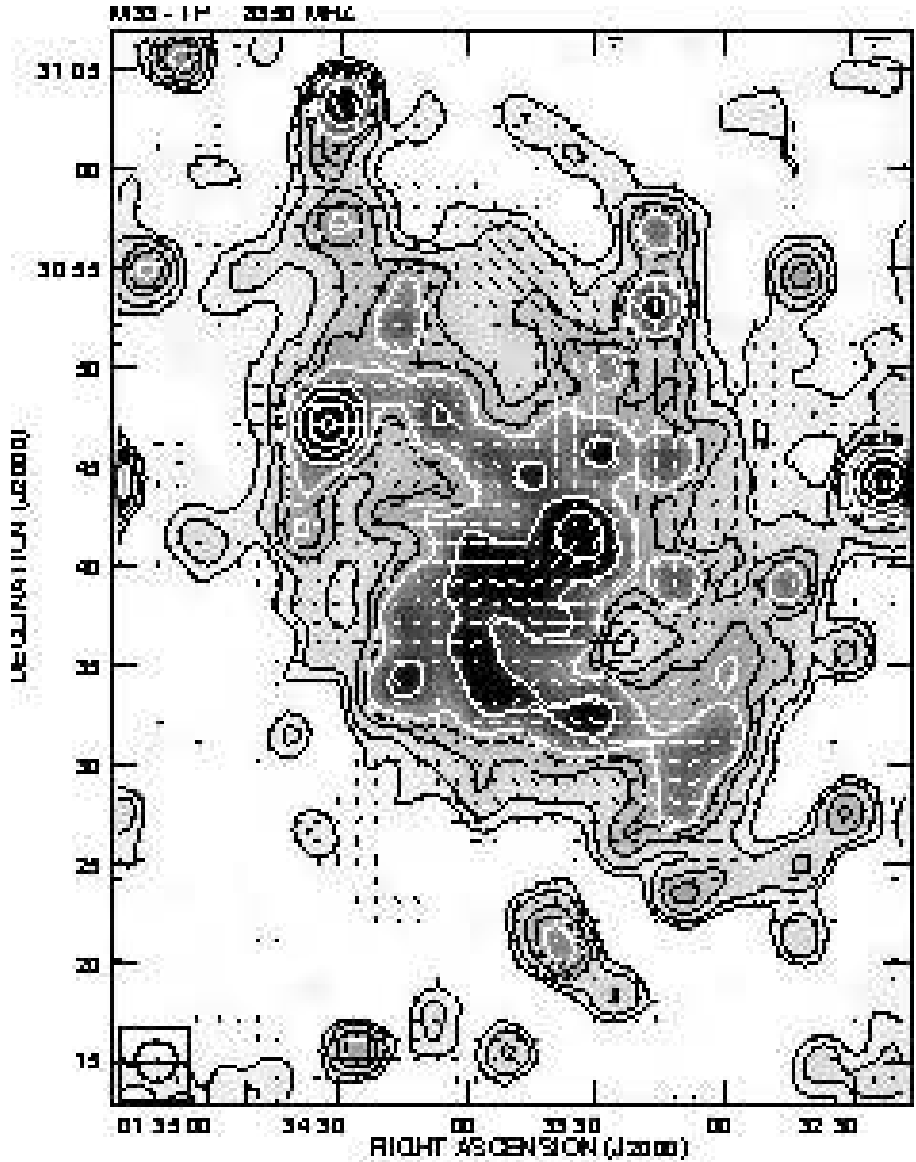}}
\caption[]{Total intensity emission from M33 at 3.6\,cm observed with the Effelsberg telescope, with apparent B-vectors   (E-vectors rotated by 90$^{\circ}$) of polarized intensity superimposed. The HPBW is 120$\arcsec$ (the beam area is shown in the left--hand corner). Contour levels are 0.5, 1, 2, 3, 4, 8, 16, 32, 64\,mJy/beam. The rms noise is 0.14\,mJy/beam in total intensity and 0.04\,mJy/beam in linear polarization. The vectors were plotted starting from 0.1\,mJy/beam (2.5\,$\sigma$). A vector length of 1$\arcmin$ represents a polarized intensity of 0.5\,mJy/beam. }
\end{figure*}

\begin{figure*}
\begin{center}
\resizebox{10cm}{!}{\includegraphics*{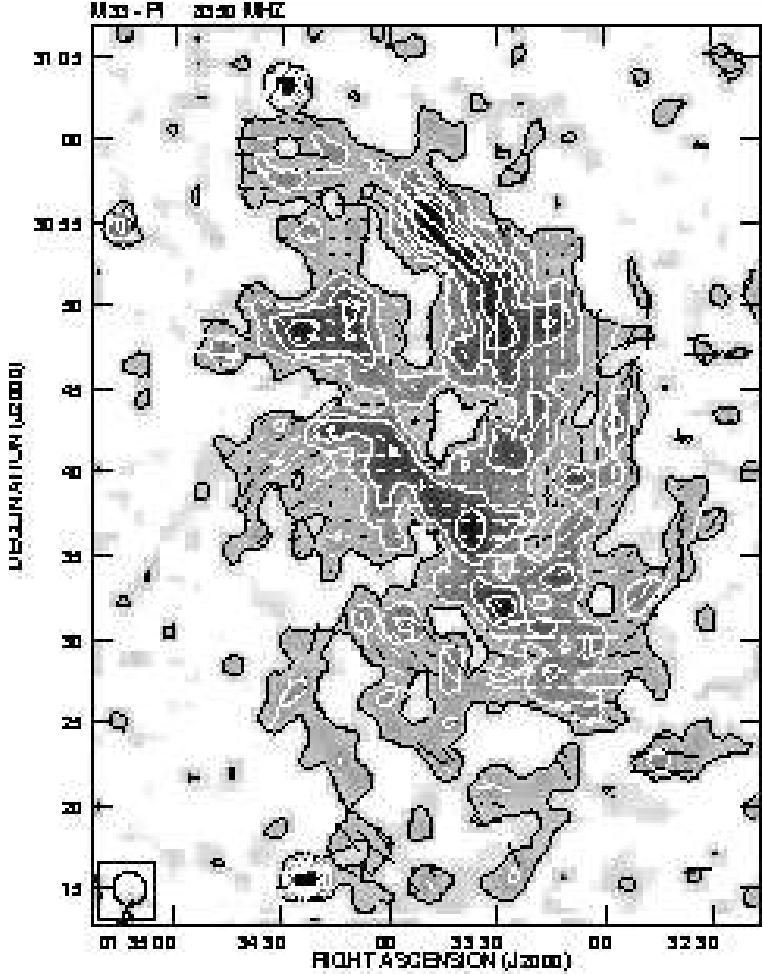}}
\caption[]{Linearly polarized emission (contours and grey scale) and degree of polarization (vector lengths) from M33 at 3.6\,cm observed with the Effelsberg telescope, smoothed to an angular resolution of 120$\arcsec$ (the beam area is shown in the left--hand corner). Position angles of the vectors show the orientation of the apparent magnetic field component perpendicular to the line of sight. Contour levels are 0.1, 0.2, 0.3, 0.4\,mJy/beam. The rms noise is 0.04\,mJy/beam. A vector length of 1$\arcmin$ represents a degree of polarization of 12.6$\%$. }
\end{center}
\end{figure*}

\begin{figure*}
\begin{center}
\resizebox{\hsize}{!}{\includegraphics*{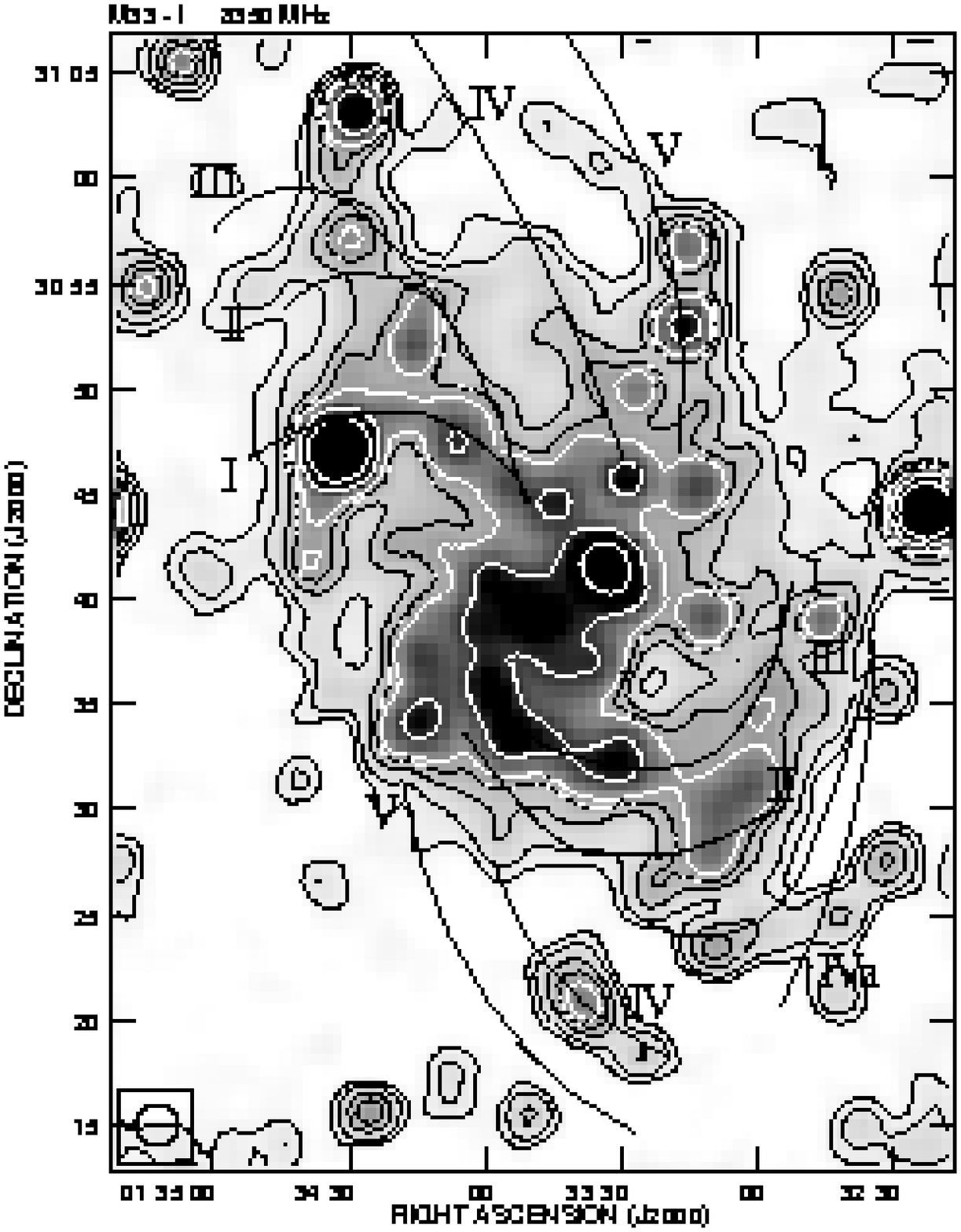}
\includegraphics*{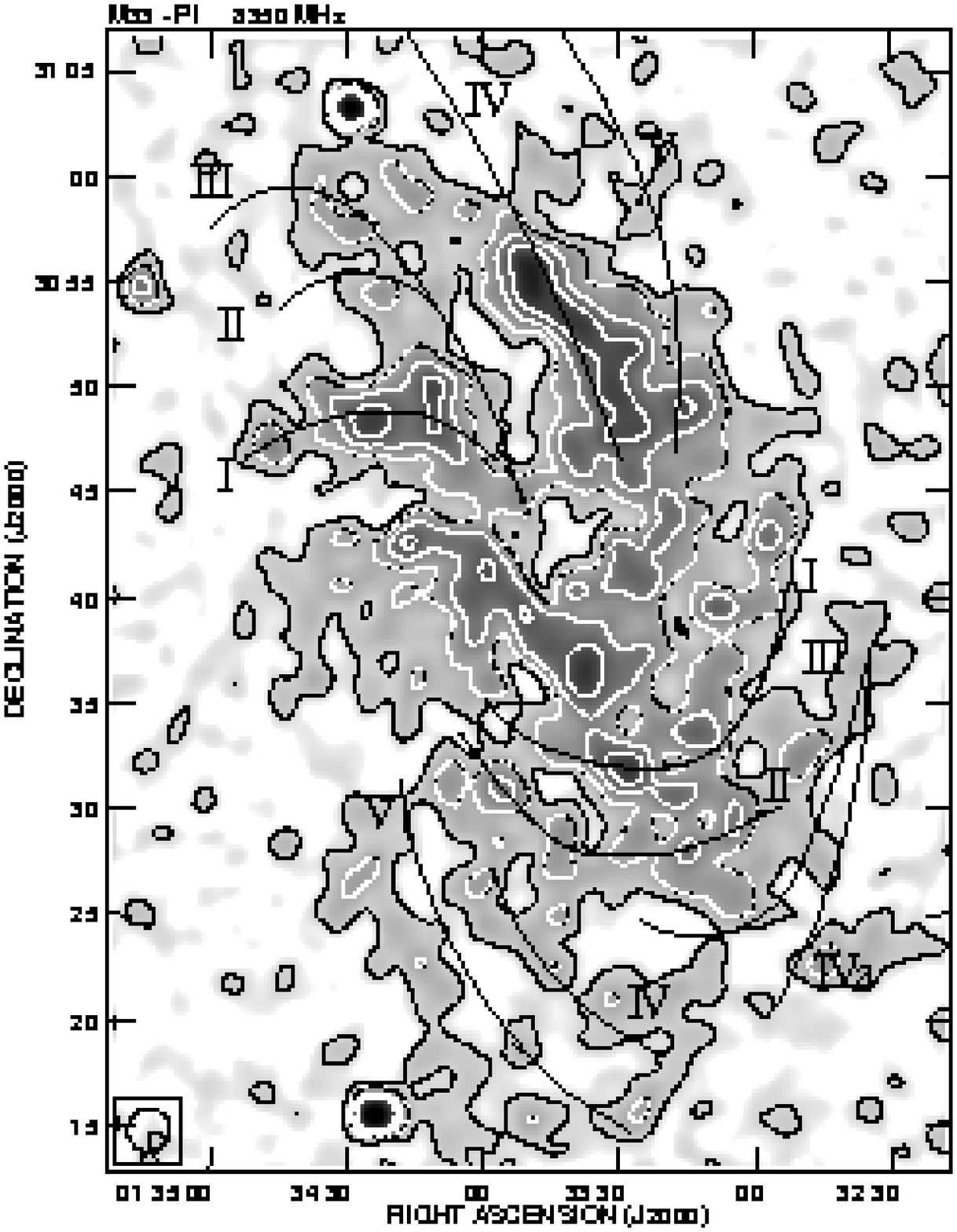}}
\caption[]{A sketch of the optical arms \citep{Sandage} overlayed on the total (left panel) and linearly polarized (right panel) intensities at 3.6\,cm. Grey scales and Contour levels are the same as in Figs. 1 and 2.}
\end{center}
\end{figure*}

\begin{figure*}
\begin{center}
\resizebox{12cm}{!}{\includegraphics*{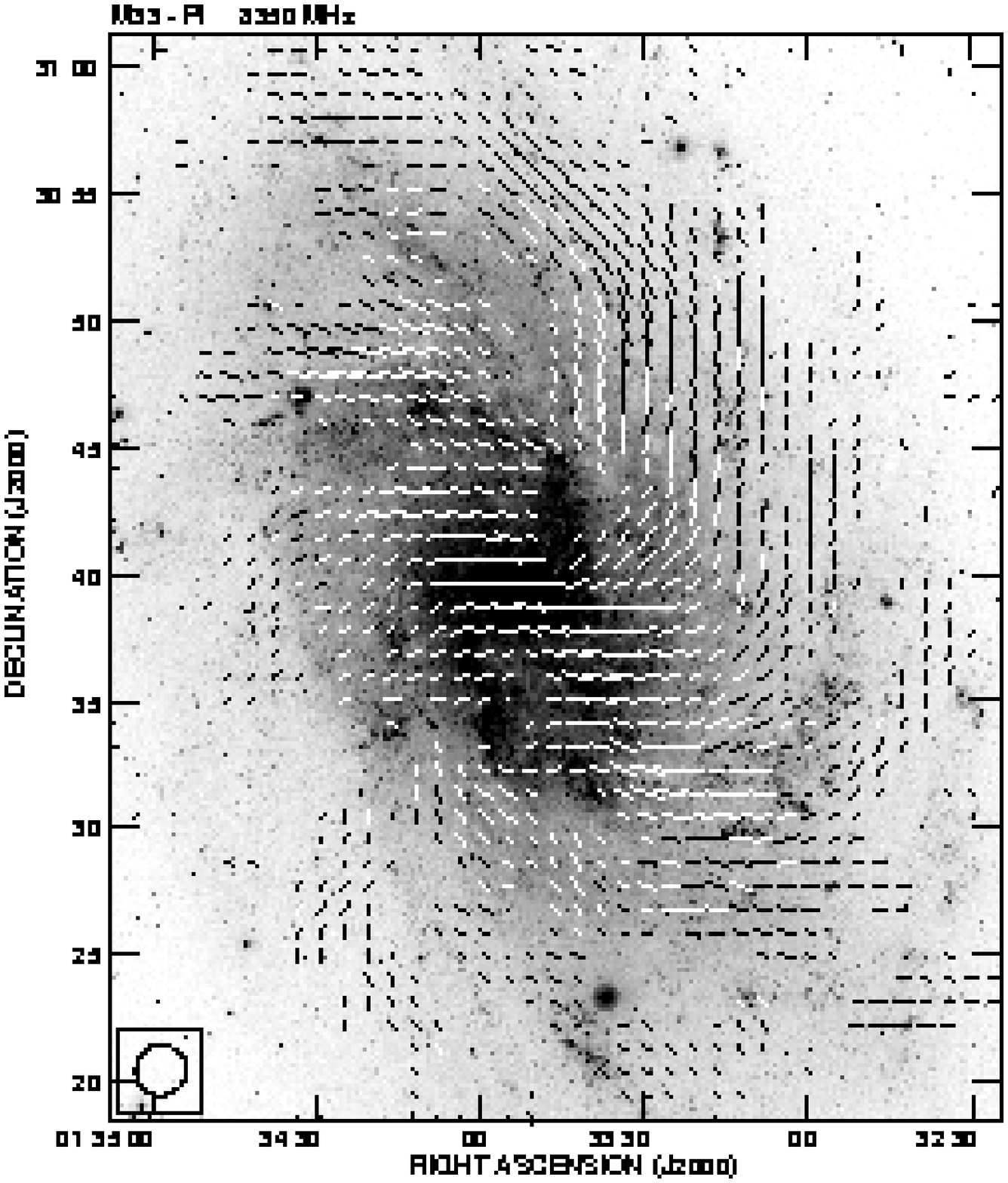}}
\caption[]{Apparent B-vectors (E-vectors rotated by 90$^{\circ}$) at 3.6\,cm superimposed on an optical image (B-band, taken from the STScI Digitized Sky Survey)  of M33. A vector length of 1$\arcmin$ represents a polarized intensity of 0.3\,mJy/beam. The vectors were plotted starting from 1.5\,$\sigma$. }
\end{center}
\end{figure*}

Observations at 6.2\,cm were carried out in the summer of 1996 using the 4850 MHz HEMT receiver installed in the secondary focus of the  100--m Effelsberg telescope. This is a 4-channel 2-beam system with cooled HEMT pre-amplifiers (see Table 2 for more information about this receiver).  M33 was scanned along its major axis and perpendicular to it  on a grid of 1$\arcmin$ with a scanning velocity of 90$\arcmin$/min. The field size is 100$\arcmin$\,$\times$\,80$\arcmin$. The number of useful coverages is 12 for Stokes I maps and 14 for each U and Q maps.  The 6.2\,cm data reduction was performed similar to that at 3.6\,cm. The r.m.s. noise after combination is $\sim$\,700\,$\mu$Jy/beam in Stokes I (total intensity) map and $\sim$\,180\,$\mu$Jy/beam in Stokes U and Q (polarization) maps. The final maps of total and polarized intensities (smoothed to 180$\arcsec$) are shown in Figs. 5 and 6, respectively.

\begin{figure*}
\begin{center}
\resizebox{11cm}{!}{\includegraphics*{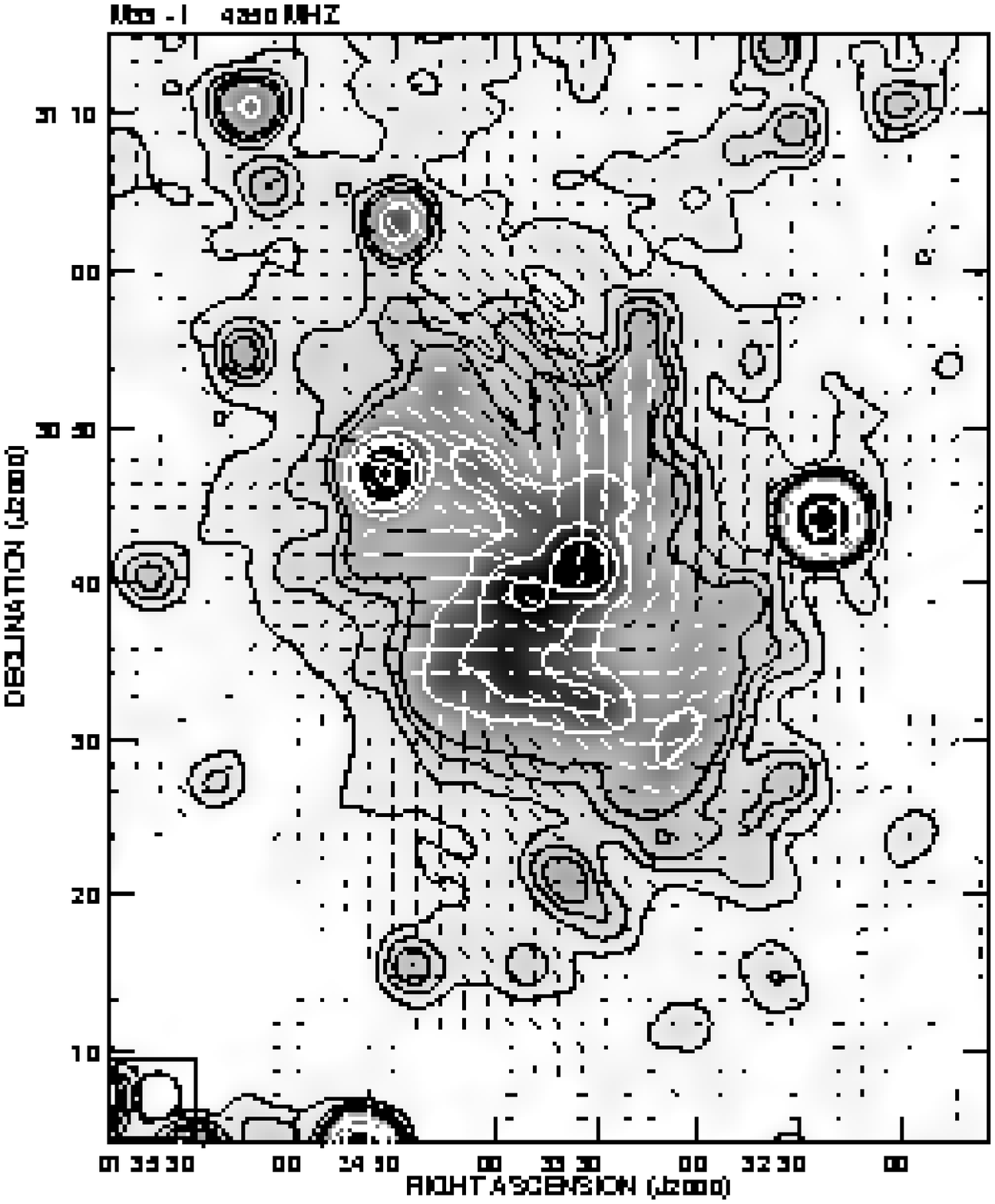}}
\caption[]{Total intensity emission from M33 at 6.2\,cm observed with the Effelsberg telescope, with an angular resolution of 180$\arcsec$ (the beam area is shown in the left--hand corner). Apparent B-vectors of the polarized intensity (E-vectors rotated by 90$^{\circ}$) are superimposed. Contour levels are 2.2, 4.4, 6.6, 8.8, 17.6, 26.4, 35.2, 70.4\,mJy/beam. The rms noise is 0.56\,mJy/beam in total intensity and 0.12\,mJy/beam in polarization. The vectors were plotted starting from 2.5\,$\sigma$. A vector length of 1.5$\arcmin$ represents a polarized intensity of 1\,mJy/beam. }
\end{center}
\end{figure*}

\begin{figure*}
\begin{center}
\resizebox{11cm}{!}{\includegraphics*{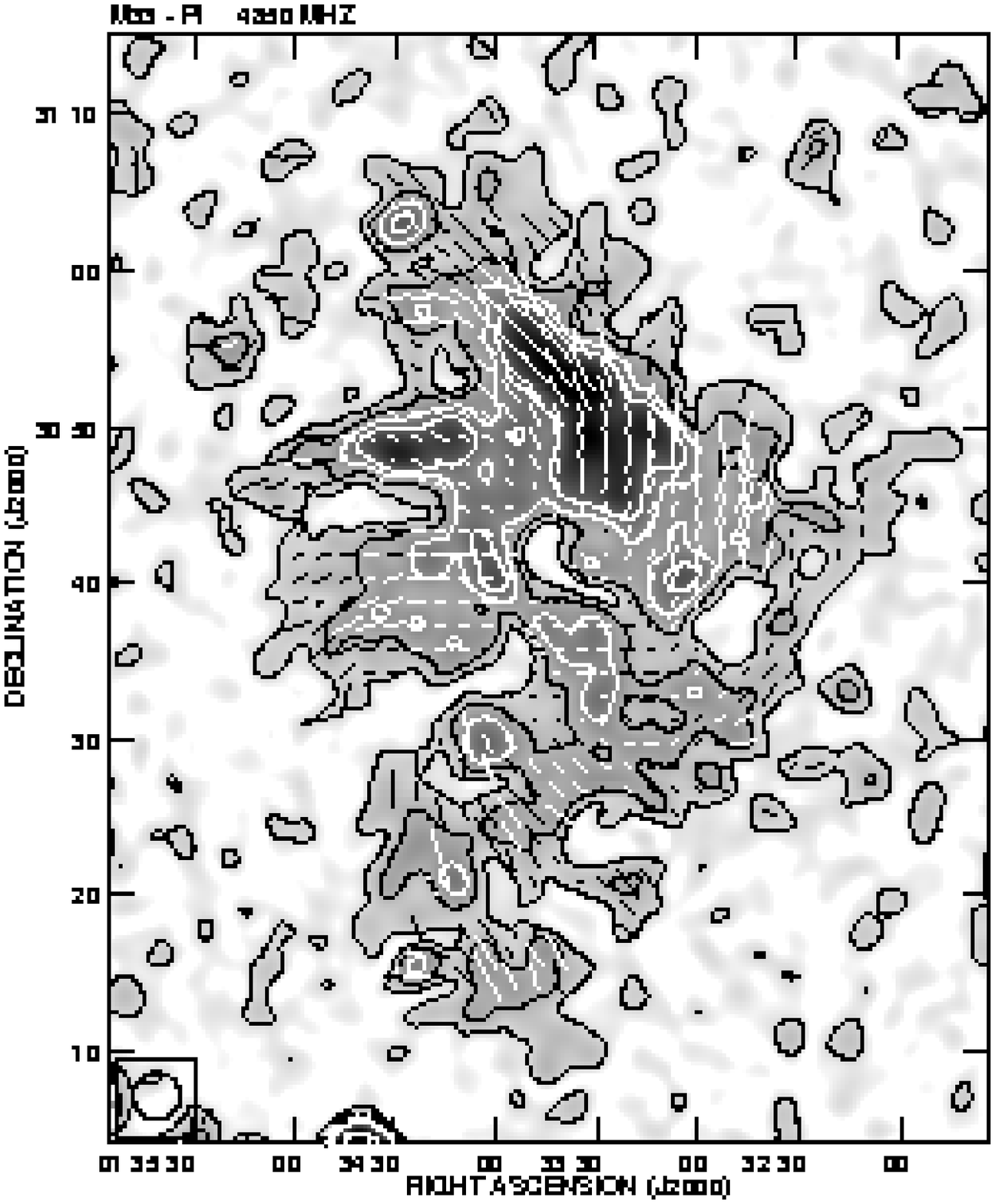}}
\caption[]{Linearly polarized emission (contours and grey scale) and degree of polarization (vector lengths) from M33 at 6.2\,cm observed with the Effelsberg 100--m dish, with an angular resolution of 180$\arcsec$ (the beam area is shown in the left--hand corner). Position angles of the vectors show the orientaiton of the magnetic field component perpendicular to the line of sight. Contour levels are 0.3, 0.6, 0.9, 1.2, 2.4\,mJy/beam area. The rms noise is 0.12\,mJy/beam. A vector length of 1.5$\arcmin$ represents a degree of polarization of 11.7$\%$.}
\end{center}
\end{figure*}

\begin{table*}
\begin{center}
\caption{Central positions of the 12 fields observed with VLA at 20\,cm.}
\begin{tabular}{ l l l l l l} 
\hline

 Pointing &   RA\,($\alpha$) &   DEC\,($\delta$) & I rms noise & U\,(Q) rms noise & HPBW\\
 $\#$   &   ($^h$ $^m$ $^s$) & ($^{\circ}$ $\arcmin$ $\arcsec$)& ($\mu$Jy/beam)& ($\mu$Jy/beam)&(arcsec$^2$)\\
          & & & &\\ 
\hline 
\hline
1  & 01 35 01.26 & +30 16 45.00 & 26 &23 & 50.7\,$\times$\,49.5 \\
2  & 01 33 51.79 & +30 16 45.00 & 23 &20 & 50.2\,$\times$\,46.1\\
3  & 01 32 42.31 & +30 16 45.00 & 27 &18& 50.6\,$\times$\,43.4 \\
4  & 01 35 01.26 & +30 31 45.00 & 26 &21 & 48.1\,$\times$\,44.0\\
5  & 01 33 51.60 & +30 31 45.00  & 28 &20& 47.8\,$\times$\,45.4\\
6  & 01 32 41.96 & +30 31 45.00  & 32 &19& 51.4\,$\times$\,42.2\\
7  & 01 35 01.26 & +30 46 45.00  & 26 &21& 50.0\,$\times$\,46.1\\
8  & 01 33 51.41 & +30 46 45.00  & 34 &21& 49.2\,$\times$\,42.4\\
9  & 01 32 41.58 & +30 46 45.00  & 23 &19& 51.1\,$\times$\,44.0\\
10 & 01 35 01.26 & +31 01 45.00  & 36 &23& 48.5\,$\times$\,45.2\\
11 & 01 33 51.21 & +31 01 45.00  & 24 &21& 51.1\,$\times$\,43.3\\
12 & 01 32 41.21 & +31 01 45.00  & 21 &19& 48.0\,$\times$\,44.0\\
\hline
\end{tabular}
\end{center}
\end{table*}

\subsection{VLA observations}

We observed M33 at 20\,cm (L--band) in two 50\,MHz bands centered at 1385 and 1465 MHz with dual circular polarization in the continuum mode of the VLA D-array. The observation dates were 06-11-05, 07-11-05, 08-11-05, 13-11-05, and 06-01-06. To cover the central $30\arcmin$\,$\times$\,$45\arcmin$ with equal sensitivity, we made 
a mosaic with 12 pointings with a spacing of half the primary beam width, 
$15\arcmin$, (3 pointings along RA and 4 pointings along DEC). The coordinates of the 12 pointings are listed in Table 2. The observations were made in cycles with 6\,min on each pointing of the mosaic. This observing cycle was repeated 17 times.  At the beginning and end of each 4--cycle, we observed 3C\,138 and 3C\,48 as the primary flux density calibrators. The antenna gains and phases were calibrated every 43\,min with the phase calibrator 0029\,+\,349.  The source 3C\,138 was also used for polarization calibration.

The standard procedures of the Astronomical Image Processing System (AIPS) were used to reduce the data. The data from each day were calibrated separately to ensure that there were no day--to day amplitude discrepancies. Then they were split into 12 single--source visibility data sets corresponding to the different pointings. After flagging, we combined the source visibility data from different days of the same source.  We self--calibrated the combined visibility data in both phase and amplitude\,$\&$\,phase.  Final images were made with IMAGR parameter Robust=0. The rms noise of different pointings in both total intensity (I) and polarization (U and Q) are listed in Table 3. We convolved all the single images to the largest clean beam of 51.4$\arcsec$ in both RA and DEC, the individual images were reprojected on a single grid centered at RA\,=\,01$^h$ 33$^m$ 51.41$^s$ and DEC\,=\,+30$^{\circ}$ 46$\arcmin$ 45.00$\arcsec$. Then, we combined the single images to a mosaic with equal weights using the task `LTESS' which also corrects for primary beam attenuation. 
\begin{figure*}
\begin{center}
\resizebox{\hsize}{!}{\includegraphics*{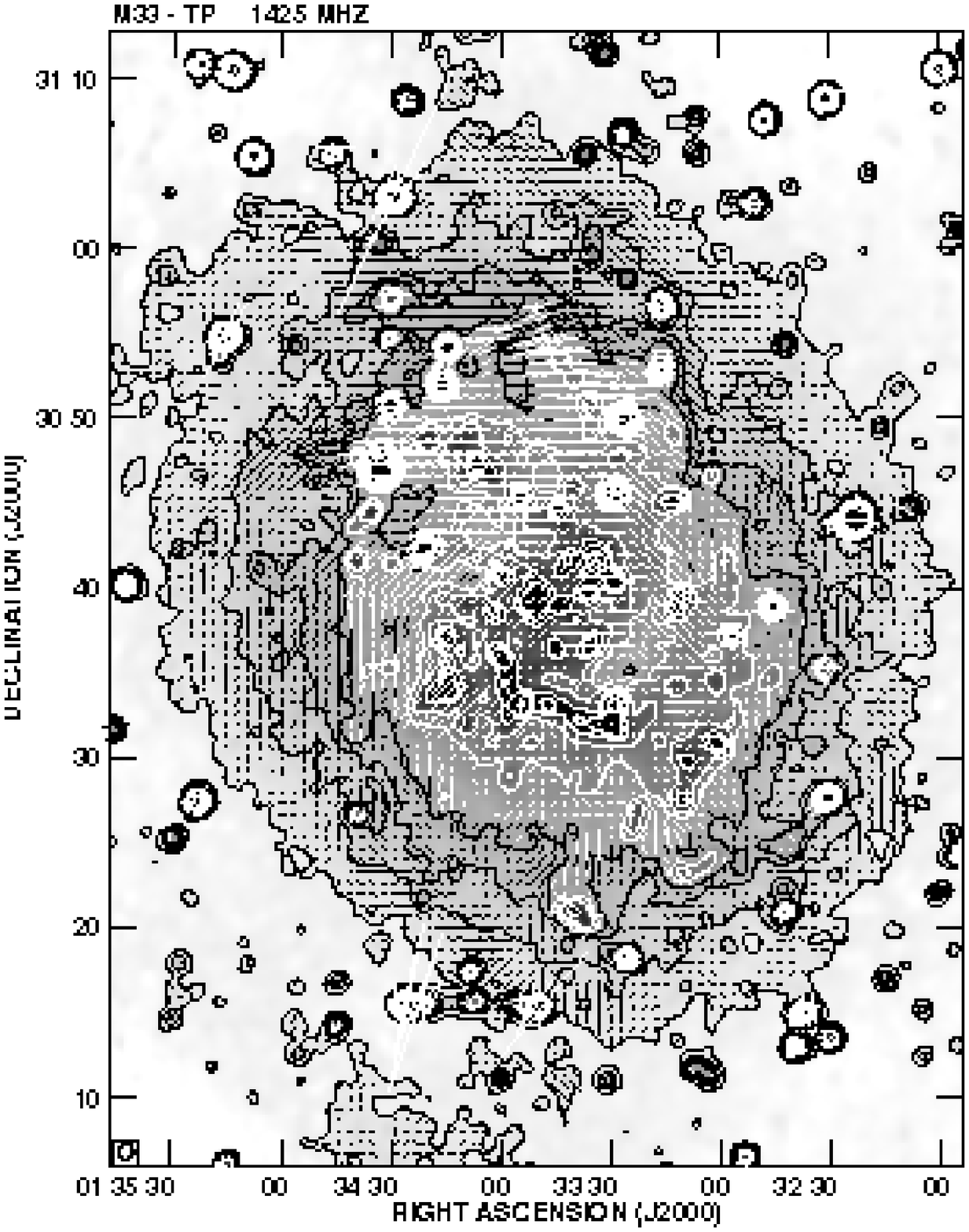}}
\caption[]{The combined VLA\,+\,Effelsberg map of the total intensity emission from M33 at 20\,cm, with apparent B-vectors of polarized intensity (E-vectors rotated by 90$^{\circ}$) superimposed. The HPBW is 51$\arcsec$ (the beam area is shown in the left--hand corner). Contour levels are 0.3, 0.6, 0.9, 1.2, 1.8, 2.4, 3.6, 4.8, 9.6, 19.2\,mJy/beam. The rms noise is 0.07\,mJy/beam in total intensity and 0.025\,mJy/beam in polarization. The vectors were plotted starting from 1.5\,$\sigma$. A vector length of 51$\arcsec$ represents a polarized intensity of 0.1\,mJy/beam. }
\end{center}
\end{figure*}

\begin{figure*}
\begin{center}
\resizebox{11cm}{!}{\includegraphics*{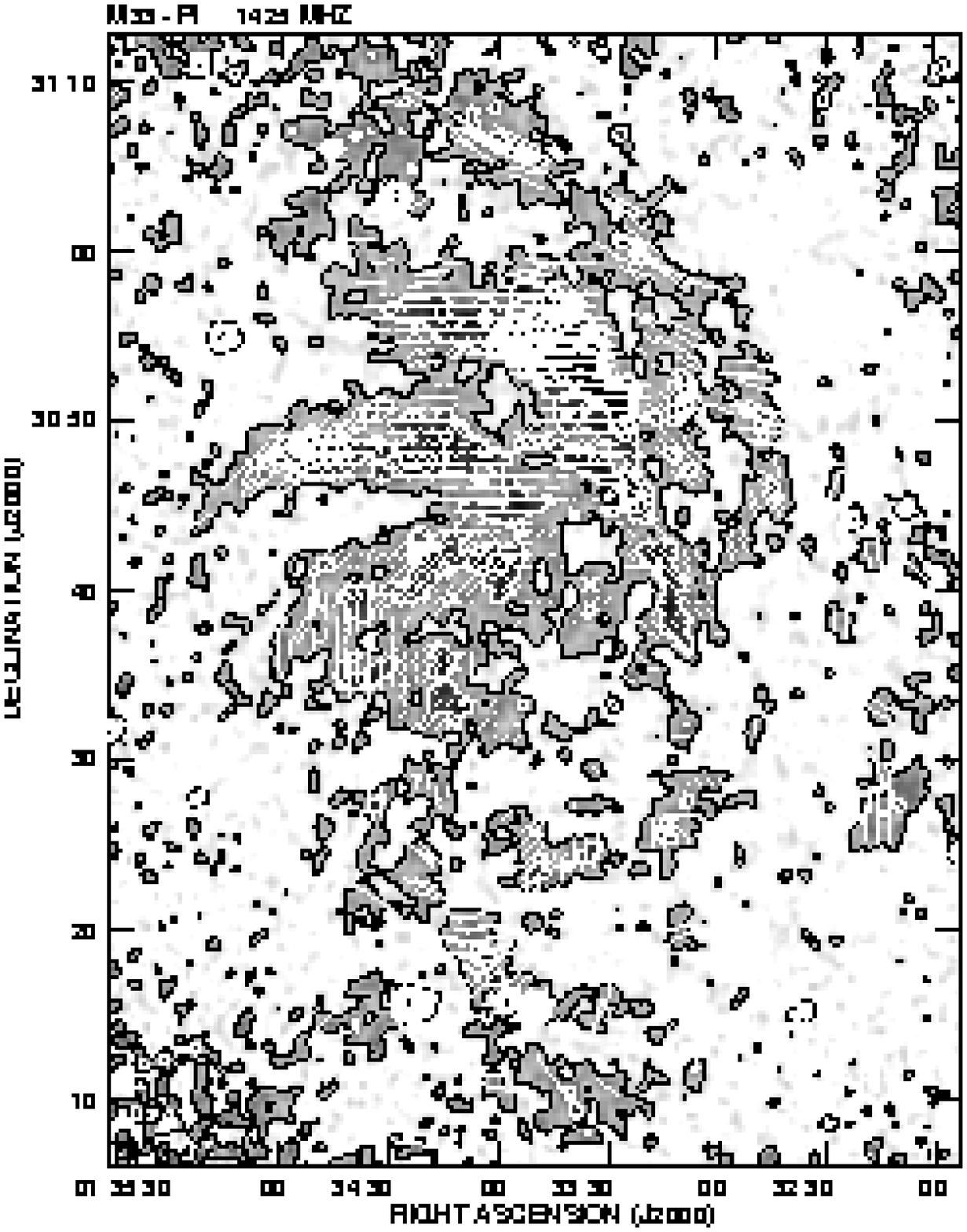}}
\caption[]{The combined VLA\,+\,Effelsberg map of the linearly polarized emission (contours and grey scale) and degree of polarization (vector lengths) from M33 at 20\,cm. The HPBW is 51$\arcsec$ (the beam area is shown in the left--hand corner). Contour levels are 0.07, 0.14, 0.21, 0.28, 0.42, 0.56, 0.84\,mJy/beam. The rms noise is 0.015\,mJy/beam. A vector length of 51$\arcsec$ represents a degree of polarization of 8.7$\%$.}
\end{center}
\end{figure*}

There is a considerable lack of flux density in the VLA maps: the integrated total flux is only $\sim$\,15$\%$ in total intensity and $\sim$\,55$\%$ in polarized intensity of those in the Effelsberg maps (determined as described in Sect.~3.3). This is not surprising because of the missing short spacings in the forms of severe attenuation of angular scales larger than $\sim \,13\arcmin$ (corresponding to the shortest spacing of $\sim$\,40\,m) and the negative bowl artifact around strong sources. 
To correct for the missing spacing, we combined the VLA map with the Effelsberg single-dish 20\,cm map \citep{Fletcher} in the uv plane. To avoid distortions  in the combined total intensity (I) map due to strong sources, we first subtracted these sources from the VLA map. The sources were identified in the Effelsberg map by comparing with the VLA map smoothed to the Effelsberg resolution. Then they were also subtracted from the Effelsberg map. The residual VLA and Effelsberg maps were combined using  the task `IMERG' with a maximum spacing of 0.236\,k$\lambda$ for the Effelsberg map and a minimum spacing of 0.170\,k$\lambda$ for the VLA map in total intensity. The corresponding maximum and minimum spacing adopted in polarizarion\,(U and Q) are 0.171\,k$\lambda$ and  0.150\,k$\lambda$, respectively.   These maximum and minimum spacing values were chosen so that the integrated flux densities of the merged maps became equal to those of the Effelsberg maps. After combination, the VLA point sources were added again.  The r.m.s. noise in the combined maps are $\sim$\,70\,$\mu$Jy/beam area in Stokes I and $\sim$\,25\,$\mu$Jy/beam area in Stokes U and Q. The resultant  total and polarized intensity images are shown in Figs. 7 and 8, respectively.

\section{Results}

\subsection{Total intensity}

Our total intensity map of M33 at 3.6\,cm  is a factor of $\sim$\,6 more sensitive than the 2.8\,cm map of \cite{Buczilowski_87}. 
The map at 3.6\,cm with the original angular resolution of 84$\arcsec$ was presented by \cite{Tabatabaei_1_07}. Even after smoothing to 120$\arcsec$, the total intensity map (Figs. 1 and 3) shows that not only the two main optical arms I\,S\footnote{We refer to the arm notation of \cite{Humphreys_80}; I\,N to V\,N for the northern and I\,S to V\,S for the southern arms.} and I\,N are well pronounced, but also the northern arms II\,N, III\,N, IV\,N, and V\,N are (at least partly) visible at 3.6\,cm. The total intensity of the extended emission is about 4\,mJy/beam in the arm I\,N which increases to 8\,mJy/beam in parts of the arm I\,S. The intensity is smaller than 3\,mJy/beam in the other arms.  The flux densities of 11 brightest HII regions at 3.6\,cm are presented in Table 4.

\begin{table}
\begin{center}
\caption{The 11 brightest HII regions at 3.6\,cm.}
\begin{tabular}{ l l l l } 
\hline
Object &   RA\,($\alpha$) &   DEC\,($\delta$) &  Flux density \\
    &   ($^h$ $^m$ $^s$) & ($^{\circ}$ $\arcmin$ $\arcsec$)& (mJy/beam)\\
\hline
NGC604 & 1 34 32.9 & 30 47 19.6& 53.7\,$\pm$\,0.4 \\
NGC595 &1 33 32.4 &30 41 50.0 & 20.7\,$\pm$\,0.7\\
IC133 & 1 33 15.3 &30 53 19.7 & 9.4\,$\pm$\,0.5\\
B690 & 1 34 06.5 &30 47 40.0 & 5.3\,$\pm$\,0.3\\
B61/62 &1 33 44.8 &30 44 50.0 & 4.9\,$\pm$\,0.5\\
IC132 & 1 33 17.1 &30 56 51.8 & 4.7\,$\pm$\,0.2\\
IC131 & 1 33 12.5 &30 45 21.5 & 4.7\,$\pm$\,0.3\\
NGC588 & 1 32 44.7 &30 39 20.6 & 4.5\,$\pm$\,0.3\\
IC142 & 1 33 56.7 &30 45 51.8 & 4.1\,$\pm$\,0.2\\
B691 & 1 34 15.4 &30 52 21.8 & 4.1\,$\pm$\,0.5\\
NGC592 & 1 33 12.6 &30 38 51.3 & 4.1\,$\pm$\,0.3\\

\hline
\end{tabular}
\end{center}
\end{table}

At 6.2\,cm, the rms noise (0.7\,mJy/beam) is improved by a factor of two with respect to the previous observations by \cite{Buczilowski_87}.   The northern arms and I\,S plus parts of the southern arms II\,S and III\,S are visible  in the second contour in Fig.~5.

With a linear resolution of $\sim$\,200\,pc (HPBW\,=\,51.4$\arcsec$) and a sensitivity of 70\,$\mu$Jy/beam, a large number of point sources are revealed in the total intensity map at 20\,cm (Fig. 7). A list of the flux densities of the radio point sources at 20\,cm at 7$\arcsec$ resolution was presented by \cite{Gordons_99}\footnote{They observed M33 with the VLA and WSRT at 1.4\,GHz,  measured the flux density of point sources, but  missed the extended emission.}. The contour levels shown in Fig. 7 starts  at the 4$\sigma$ level. The strongest extended emission is found in the arm I\,S with a flux density of $\geq$\,3.6\,mJy/beam, followed by parts of the arm I\,N and the extended central region of the galaxy. At 20\,cm, the extended emission is detected up to larger radii and is more radially symmetric than at 3.6\,cm and  6.2\,cm.

\subsection{Distribution of polarized emission}

The polarized emission at 3.6\,cm is shown in Fig.~2 in grey scale and with contours.  A sketch of the optical spiral arms by \cite{Sandage} superimposed on the polarized emission is shown in Fig.~3. Strong polarized intensities (more than 0.4\,mJy/beam) are found in between and parts of the northern optical arms IV\,N and I\,N (see Fig. 3) and also in central regions of the galaxy. It seems that the polarized emisson, as a whole, obeys a spiral pattern structure. Whether this structure mimics the optical arm structure or not can be seen in Fig. 3, where a sketch of optical arms \citep{Sandage} is superimposed on the polarized intensity map. The polarized emission is found partly in the optical arms and partly in between the arms. In the other words, there is no clear  correlation between the optical arms and polarization arm-like structure. The vectors in Fig. 1 (and also 2) show the intrinsic orientation of the ordered magnetic field in the plane of the sky (PI\,$\sim$\,B$_{reg \perp}^2$), as Faraday rotation is not significant at 3.6\,cm. A comparison between the orientation of the ordered magnetic field and optical structure is shown in Fig.~4.   Again, it is seen that the ordered magnetic field traces out some sort of spiral structure which does not coincide well with the optical spiral arms, although there is a general similarity in the orientations. 
The strongest arm like filament of the magnetic field  exists in between the northern optical arms IV\,N and V\,N, but also covers parts of these two arms. Generally, in some regions the ordered magnetic field alignes parts of the optical arms (like parts of the arms I\,N and V\,N), or is driven to the optical interarms, or locally distorted in orientation. The latter is mostly seen in the southern part of the galaxy. 

The vector lengths in Fig. 2 represent the degree of polarization, P. Apart from in the northern arm IV\,N, P is not only high where the polarized intensity is strong: the southern optical arms I\,S, II\,S also show high P (larger than 20$\%$). The  highest degree of polarization, $\sim$\,40$\%$, is found near NGC604 at RA\,=\,01$^h$ 34$^m$ 49.67$^s$ and DEC\,=\,+30$^{\circ}$ 47$\arcmin$ 19.49$\arcsec$ (Fig. 3). 
However, it is less than 5$\%$ within the giant HII regions (eg. NGC604 and NGC595).   This plus a weaker ordered magnetic field (and low P) in the central southern arm (I\,S, the strongest arm in radio, IR and optical bands) indicates that high star forming activity disturbs locally the regular magnetic field and causes the patchy distribution of the degree of polarization within the arms.

At the angular resolution of 180$\arcsec$ at 6.2\,cm,  the magnetic arm like filaments are smoother and not well distinguishable  (Fig. 6). The strong polarized emission (more than 1.2\,mJy/beam) again occurs in parts of the arms IV\,N and I\,N, but in contrast to that at 3.6\,cm, not in the central regions of the galaxy (Fig. 5). The high P at 6.2\,cm ($>$ 20$\%$) has almost the same distribution as at 3.6\,cm.
In the arm IV\,N, the maximum P is about 30$\%$, similar to that at 3.6\,cm.

Fig. 7 shows, at a linear resolution of $\sim$\,200\,pc,  how the apparent magnetic field vectors (E+90$^{\circ}$) at 20\,cm are distributed in the plane of the sky. At this wavelength, the vectors are strongly Faraday rotated and therefore are not oriented along the optical arms.  The polarized signal is strongest in the northern part of the galaxy, where the ordering in a spiral like pattern is very clear (Fig.~8), while the lack of polarized emission in the southern part suggests stronger Faraday depolarization in the southern than in the northern part of M33 (see Sect. 3.6). In the northern arms, the degree of polarization P is larger than 20$\%$,  with a maximum of 30$\%$  on  the arm IV\,N and also in a region close to NGC604 in I\,N.

\subsection{Integrated flux densities}

After subtraction of the background sources (14 sources whose 20\,cm flux densities were $\geq$\,5\,mJy/beam), we integrated the flux densities in rings  in the plane of the galaxy around the center out to a radius of 8.5\,kpc (beyond the area shown in Figs. 1 to 8)(see Table 1). The integrated flux density in total intensity, S$_{T}$, and in linear polarization, S$_{PI}$,  at different wavelengths are given in Table 5. The corresponding absolute total flux density errors were obtained taking into account the systematic error due to the baselevel uncertainty. 

\begin{table*}
\begin{center}
\caption{Integrated flux densities for $R$\,$<$\,8.5\,kpc (35$\arcmin$).}
\begin{tabular}{ l l l l l l l } 
\hline
$\lambda$ & HPBW & S$_{T}$ &  S$_{PI}$  & $\bar{{\rm P}}$ & S$_{PI}$(180$\arcsec$) & $\bar{{\rm P}}$(180$\arcsec$)\\
(cm) & (arcsec$^2$)& (mJy) & (mJy) & $\%$ & (mJy) &$\%$ \\
\hline 
\hline
3.6 & 120 & 779\,$\pm$\,66    & 48\,$\pm$\,5 & 6.1\,$\pm$\,0.8 & 40\,$\pm$\,4 & 5.1\,$\pm$\,0.7 \\
6.2 & 180 & 1284\,$\pm$\,135   &  89\,$\pm$\,4 & 6.9\,$\pm$\,0.8 & 89\,$\pm$\,4 & 6.9\,$\pm$\,0.8 \\
20 (combined) & 51 & 2768\,$\pm$\,63 & 167\,$\pm$\,25 &  6.0\,$\pm$\,0.9 & 123\,$\pm$\,3 & 4.4\,$\pm$\,0.1 \\
\hline
\end{tabular}
\end{center}
\end{table*}

The average degree of polarization ($\bar{{\rm P}}$), which is the ratio of the integrated flux density in polarized intensity and the integrated flux density in total intensity, is obtained at different wavelengths (Table 5). For a consistent comparison, we also obtained the average degree of polarization after smoothing the 3.6\,cm and 20\,cm maps to the beam width of  180$\arcsec$, $\bar{{\rm P}}$(180$\arcsec$).
At 3.6\,cm and 20\,cm, $\bar{{\rm P}}$(180$\arcsec$) is smaller than $\bar{{\rm P}}$ due to beam depolarization. $\bar{{\rm P}}$(180$\arcsec$) is almost the same at  3.6\,cm and 6.2\,cm, indicating  negligible global Faraday depolarization between these two wavelengths. 

The integrated total flux densities in total intensity obtained by \cite{Buczilowski_88} of 1100\,$\pm$\,167\,mJy at 6.3\,cm and  2990\,$\pm$\,440\,mJy at 20\,cm out to a radius of 12\,kpc agree with our values of 1300\,$\pm$\,140\,mJy at 6.3\,cm and 3164\,$\pm$\,88\,mJy at 20\,cm within the errors. Our obtained value at 6.2\,cm is even closer to that obtained by  \cite{Berkhuijsen_83} (1300\,$\pm$\,200\,mJy).
The integrated total flux density in total intensity found by \cite{Buczilowski_88} at 2.8\,cm is significantly smaller than that we found at 3.6\,cm because the galaxy size is much larger than the maximum spacing of the 2.8\,cm multibeam system  \citep{Emerson}.

\subsection{Distribution of total spectral index}
\begin{figure}
\resizebox{7.3cm}{!}{\includegraphics*{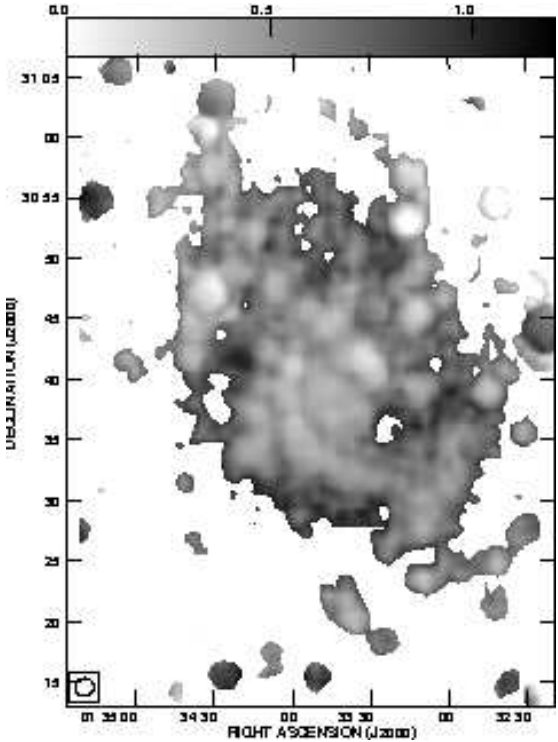}}
\caption[]{Total spectral index map of M33 as determined between the two wavelengths 3.6 and 20\,cm. The angular resolution is 90$\arcsec$. For the total intensity maps, 3$\sigma$ level was considered to obtain this spectral index map.}
\end{figure}

Because of their large frequency range, the total intensities at 20\,cm and 3.6\,cm lead to a reliable spectral index map. After smoothing  to the angular resolution of 90$\arcsec$, the total intensity maps at 20\,cm and 3.6\,cm were restricted to a common size and interpolated onto the same grid. 
The spectral index, $\alpha$ (I\,$\sim$\,$\nu^{-\alpha}$), was only computed for pixels with a flux density of at least three times the rms noise $\sigma$ at both frequencies. In the resultant spectral index map (Fig. 9), the spiral arms and star forming regions are well visible as flat spectrum regions. Steeper spectra occur in the outer regions and in between arms. Fig. 10  shows that the total spectral index has a mean value of 0.7.  The error in the total spectral index varies from 0.01 in the star forming regions to 0.23 in the outer parts. 

The HII complexes IC\,133, NGC\,604, and NGC\,595 have spectral indices $<$\,0.1 and even negative values, indicating  optically thick conditions in these HII regions.  \cite{Johnson} found several optically thick HII regions (at 6\,cm and 20\,cm) with the two brightest ones in IC\,133. 

The most inverted spectrum is due to a source at RA\,=\,01$^h$ 32$^m$ 42.06$\pm$0.14$^s$ and DEC\,=\,+30$^{\circ}$ 54$\arcmin$ 24.8$\pm$1.8$\arcsec$ which has a spectral index of $-0.6 \pm$0.1 (I$_{3.6 cm}$ = 3.6\,$\pm$\,0.2 and I$_{20 cm}$ = 1.24\,$\pm$\,0.07). This source, which does not exist in the catalogues of radio sources and HII regions of M33 \citep[e.g. in ][]{Gordons_99}, could be a variable radio source or a giant optically thick HII region behind a high concentration of dust. However, the latter possibility is unlikely since no IR emission is observed from that region. 

We obtained the radial distribution of the spectral index $\alpha$ using the integrated total flux densities calculated in rings with a width of 0.5\,kpc in the galactic plane. Fig. 11 presents the radial profile of the spectral index obtained from pairs of the  integrated flux densities at 3.6\,cm, 6.2\,cm, and 20\,cm  considering one $\sigma$ level. There is a general increase in $\alpha$ from the center to a radius of about 7\,kpc in all three profiles. \cite{Buczilowski_88} explained this increase as a general decrease of the thermal fraction with radius. However, with our high resolution study we detect structures not seen by \cite{Buczilowski_88}\,:  a prominant bump around $R$\,=\,3\,kpc and a steeper increase in $\alpha$ beyond $R$\,=\,4\,kpc. This indicates that the thermal fraction decreases faster in regions at $R$\,$>$\,4\,kpc than at smaller radii. The bump is located in the ring $2.5<$\,$R$\,$<3$\,kpc dominated by interarm regions which are expected to be weak in thermal emission.  

Assuming that the radial increase of the spectral index is related to a decrease in the thermal fraction, and considering that the thermal emission contributes more to the radio continuum emission at smaller wavelengths, one expects more variations (or faster radial increase) of the spectral index obtained at small wavelengths.   
This is indeed visible in Fig.~11\,: $\alpha(3.6,6.2)$ increases faster than $\alpha(6.2,20)$ and  $\alpha(3.6,20)$ at $R$\,$>$\,4\,kpc. It is also seen that the larger the wavelengths, the smaller the spectral index (or flatter energy spectrum at larger wavelengths). This is surprising as one expects that the spectrum of the radio continuum emission becomes steeper at larger wavelengths because of the synchrotron emission. This could hint to the non-thermal energy losses at larger wavelengths. The reason may be the same as of the significant flattening of the integrated spectrum of M33 at frequencies lower than 900\,MHz\,: free-free absorption of the non-thermal emission by a cool ($<$\,1000 K) ionized gas  \citep{Israel_92}. The presence of the optically thick HII regions \citep{Johnson} could also  enhance this trend.

\begin{figure}
\resizebox{\hsize}{!}{\includegraphics*{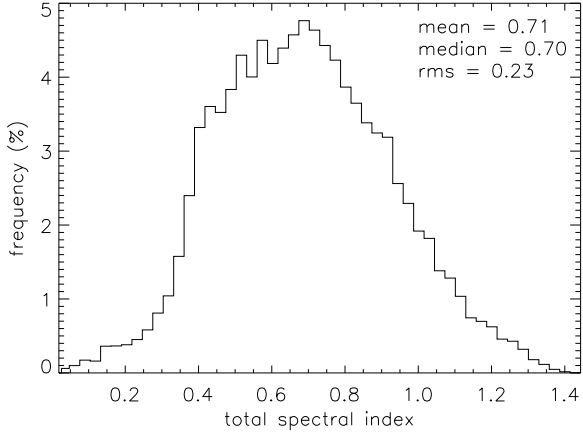}}
\caption[]{Histogram of  the spectral index distribution determined from total intensities}. 
\end{figure}

\begin{figure}
\resizebox{\hsize}{!}{\includegraphics*{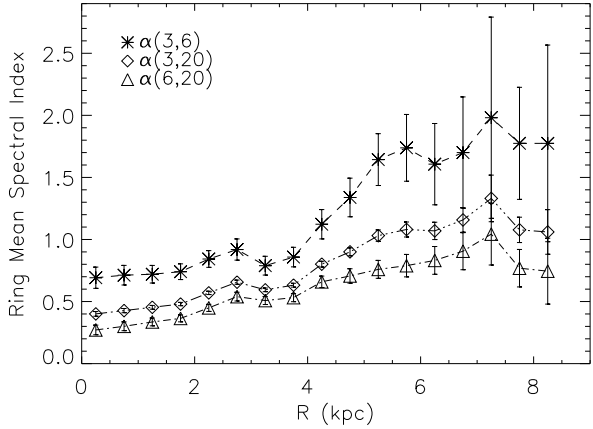}}
\caption[]{Total spectral index averaged in rings of 0.5\,kpc width in the galactic plane versus the radial distance from the center of M33.  For the total intensity maps, one $\sigma$ level was considered before integration. }
\end{figure}

\begin{table}
\begin{center}
\caption{Integrated total flux densities from Effelsberg surveys ($R<50\arcmin$).}
\begin{tabular}{ l l l } 
\hline
$\lambda$\,(cm) & $\nu$\,(MHz) & S\,(mJy) \\
\hline 
\hline
35.6 & 842 & 5377\,$\pm$\,1217$^{1}$ \\
21.1 & 1420& 2990\,$\pm$\,440$^{1}$\\
20\,(21) & 1425& 2768\,$\pm$\,63$^{2}$\\
17.4 & 1720& 2714\,$\pm$\,254$^{1}$\\
11.1 & 2702& 1683\,$\pm$\,168$^{1}$\\
6.3 &4750 &1100\,$\pm$\,167$^{1}$\\
6.2 &4850& 1300\,$\pm$\,135$^{2}$\\
3.6 & 8350 &779\,$\pm$\,66$^{2}$\\
2.8 & 10700 & 482\,$\pm$\,150$^{1}$\\
\hline
\noalign {\medskip}
\multicolumn{2}{l}{$^{1}$ \cite{Buczilowski_88}}\\
\multicolumn{2}{l}{$^{2}$ this paper }\\
\end{tabular}
\end{center}
\end{table}

Away from star forming regions  and in the outer parts of the galaxy, the thermal fraction is negligible and the observed spectral index $\alpha$ can be taken as the non-thermal spectral index, $\alpha_{n}$. From Fig. 9, we obtain $\alpha_{n} \simeq 1.0 \pm 0.1$, which is equal to that found in M31 by \cite{Berkhuijsen_03}. Assuming that the non-thermal spectral index is constant across M33, we find average thermal fractions of $0.52 \pm 0.15 $, $0.46 \pm 0.12$, and $0.18 \pm 0.06 $ at 3.6\,cm, 6.2\,cm, and 20\,cm, respectively \citep[following the method by ][]{Klein_84}. How these values change if the  non-thermal spectral index is not assumed to be constant across the galaxy will be discussed in paper II (Tabatabaei et al. in prep.).

\subsection{Integrated radio continuum spectrum}

The integrated flux densities for $R$\,$<$\,50$\arcmin$\,(12\,kpc) from Effelsberg surveys at 35.6, 21.1, 17.4, 11.1, 6.3, and 2.8\,cm were presented by \cite{Buczilowski_88}.  They obtained a spectrum with a spectral index of 0.91\,$\pm$\,0.13, steeper than found in many other spiral galaxies. 

We rederived the integrated spectrum of M33 using three more data points from the new surveys.
We calculated the integrated  flux densities for the 3.6, 6.2, and 20\,cm maps for the same area of $R$\,$<$\,50$\arcmin$. The 3.6\,cm and 2.8\,cm maps were integrated up to $R$\,=\,44$\arcmin$ and $R$\,=\,22.5$\arcmin$, respectively, providing a lower limit to the total flux density. However, extrapolated values up to $R$\,=\,50$\arcmin$ show that the missing flux densities are within errors. Our values together with those of \cite{Buczilowski_88} (see Table~6) lead to an integrated spectrum (Fig.~11) with a spectral index of 0.72\,$\pm$\,0.04, which is in agreement with the mean spectral index obtained  in Sect.~3.4  using a different method.

\begin{figure}
\resizebox{8cm}{!}{\includegraphics*{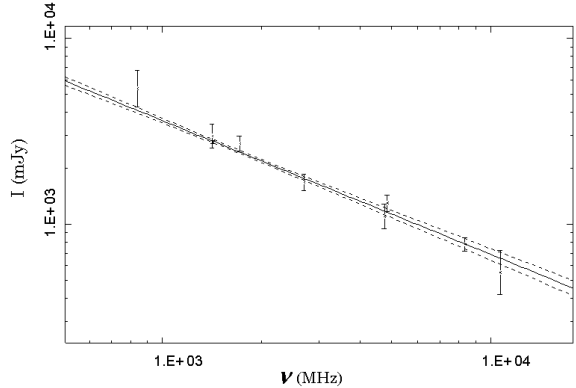}}
\caption[]{Radio spectrum of the integrated emission from M33 using the data listed in Table 6. }
\end{figure}

\begin{table}
\begin{center}
\caption{Exponential scale lengths.}
\begin{tabular}{ l l l } 
\hline
$\lambda$ & $l_{R < 4 {\rm kpc}}$\,(kpc) & $l_{R > 4 {\rm kpc}}$\,(kpc)\\
\hline 
\hline
20\,cm & 5.44\,$\pm$\,0.27  & 2.74\,$\pm$\,0.10\\
6.2\,cm & 3.73\,$\pm$\,0.21 & 2.35\,$\pm$\,0.14\\
3.6\,cm& 3.48\,$\pm$\,0.53   & 1.29\,$\pm$\,0.03 \\
24\,$\mu$m& 2.07\,$\pm$\,0.23   & 1.35\,$\pm$\,0.09 \\
70\,$\mu$m& 2.04\,$\pm$\,0.19   & 1.85\,$\pm$\,0.16 \\
160\,$\mu$m& 2.51\,$\pm$\,0.19   & 1.63\,$\pm$\,0.09 \\
\hline
\end{tabular}
\end{center}
\end{table}

\begin{figure}
\resizebox{\hsize}{!}{\includegraphics*{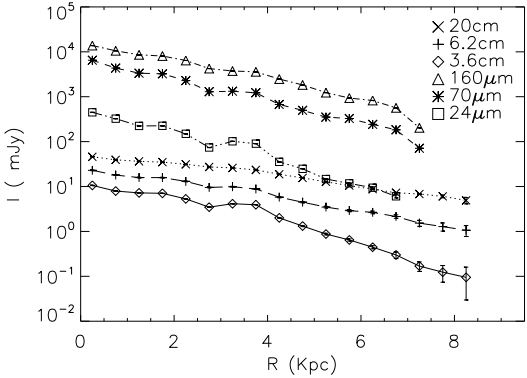}}
\caption[]{Radial profiles of total intensity (surface brightness) at radio and IR wavelengths. }
\end{figure}

\begin{table*}
\caption{Northern and southern integrated flux densities in total and polarized intensity plus degree of polarization. 
}
\begin{tabular}{ l l l l l l l l l l } 
\hline
$\lambda$ & S$_{T}^{N}$ & S$_{T}^{S}$ & S$_{T}^{N}$/S$_{T}^{S}$ & S$_{PI}^{N}$& S$_{PI}^{S}$ & S$_{PI}^{N}$/S$_{PI}^{S}$ &$\bar{{\rm P}}^{N}$ &$\bar{{\rm P}}^{S}$&$\bar{{\rm P}}^{N}$/$\bar{{\rm P}}^{S}$\\
\hline 
\hline
cm & (mJy) & (mJy) & & (mJy) & (mJy) & & $\%$& $\%$& \\
\hline
3.6 & 412\,$\pm$\,18  & 366\,$\pm$\,21 & 1.12\,$\pm$\,0.08 & 20\,$\pm$\,2 & 20\,$\pm$\,2 &1.00\,$\pm$\,0.14 & 4.8\,$\pm$\,0.4 & 5.5\,$\pm$\,0.6 & 0.9\,$\pm$\,0.1 \\
6.2 & 631\,$\pm$\,63   &  653\,$\pm$\,70 & 0.97\,$\pm$\,0.14 & 53\,$\pm$\,4 & 36\,$\pm$\,2 & 1.47\,$\pm$\,0.14& 8.4\,$\pm$\,1.0& 5.5\,$\pm$\,0.6 &1.5\,$\pm$\,0.2  \\
20 & 1268\,$\pm$\,73 & 1435\,$\pm$\,83 & 0.88\,$\pm$\,0.07& 81\,$\pm$\,5 & 42\,$\pm$\,3 & 1.93\,$\pm$\,0.18& 6.4\,$\pm$\,1.1& 2.9\,$\pm$\,0.3 & 2.2\,$\pm$\,0.4 \\

\hline
\end{tabular}
\end{table*}

\subsection{Exponential scale lengths}
Fig. 13 shows the  radial distributions of the total radio and IR (Spitzer 24, 70, and 160\,$\mu$m) intensities  in M33 with galactocentric radius. There is a break in the profiles near $R$\,=\,4\,kpc, which is best visible at 3.6\,cm and 24\,$\mu$m. All the profiles can be well fitted by two exponential functions of the form 

\begin{equation}
I(R)=\left \{ \begin{array}{ll}
{I_{0}exp(-R/l_{R < 4 {\rm kpc}})}  & \,\,\,\,\,\,\,R < 4\,{\rm kpc} \\ 
I_{4}exp(-R/l_{R > 4 {\rm kpc}}) & \,\,\,\,\,\,\, R \geq 4\,{\rm kpc} ,\\ %
\end{array} \right.
\end{equation}

where, $I_0$ and $I_4$ are the intensity at $R=0$ and $R=4$\,kpc, respectively, and $l$ is the exponential scale length.

We derived the exponential scale lengths of the radio emission at the three wavelengths along with those  of the IR emission (Table 7). Generally, the scale lengths are  larger in regions at $R$\,$<$\,4\,kpc than at $R$\,$>$\,4\,kpc. This indicates that the surface brightness decreases faster with increasing radius at $R$\,$>$\,4\,kpc at all wavelengths.  Since the 24\,$\mu$m emission \citep[and also 3.6\,cm emission, ][]{Tabatabaei_1_07} traces the star forming regions, this confirms that the main star forming regions in M33 are spread over an area with a radius of $R$\,$<$\,4\,kpc without a dominant `nuclear concentration' \citep[different from the situation in NGC6946, ][]{Walsh}. 

Table 7 also shows that the scale length increases with increasing wavelength in the radio regime. This can be explained by the fact that
at longer wavelengths the non-thermal emission becomes more dominant, and  the influence of  radial diffusion of cosmic ray electrons becomes more important, leading to larger scale lengths.

\subsection{North--south asymmetry}

Fig. 7 shows that the polarized emission at 20\,cm is much weaker in the southern than in the northern half. To investigate this asymmetry, we calculated the integrated total flux density in both total and polarized intensities separately over each half (Table 8). The average degrees of polarization are also given in Table 8. At all three radio wavelengths, the beam width is 180$\arcsec$. The ratio of the northern to the southern integrated total flux density in polarized intensity, S$_{PI}^{N}$/S$_{PI}^{S}$, and also the ratio of the average degree of polarization, $\bar{{\rm P}}^{N}$/ $\bar{{\rm P}}^{S}$, increases with wavelength.  S$_{T}^{N}$/S$_{T}^{S}$, does not change from 3.6\,cm to 6.2\,cm and from 6.2\,cm to 20\,cm within the errors, but, it decreases slightly from 3.6\,cm to 20\,cm.  
A wavelength-dependent asymmetry in the polarized intensity and average degree of polarization was also found in other galaxies e.g. NGC\,6946 (Beck, submitted to A$\&$A).  The interpretation is that there is an asymmetry in Faraday depolarization along the line of sight that may be caused in several ways:\\
- asymmetry in the distribution of the thermal emission, \\
- asymmetry in the uniformity of the magnetic field strength,\\
- special large-scale structure of the regular magnetic field, like e.g. a large-scale axisymmetric helical magnetic field \citep{Urbanik}.\\
We will investigate these possibilities after the separation of thermal/non-thermal components in a forthcoming paper (Tabatabaei et al. in prep.). 

\section{Discussion and conclusions}

Radio observations with high resolution and sensitivity enabled us to obtain much more information about the distribution of the linearly polarized emission in M33 than in previous studies, especially, at high frequencies where the Faraday effects are negligible. Using the new data at 3.6\,cm,  we found that the magnetic field is well ordered in a spiral structure with the general orientation of the optical spiral arms. No clear ordered magnetic field--optical arm anti-correlation, as observed in other galaxies like NGC6946 \citep{Beck_96}, M81 \citep{Krause_2_89}, and in parts of IC342 \citep{Krause_1_89,Krause_93}, is found in M33. Detection of high degrees of polarization (larger than 20$\%$), not only in the northern arms but also in the central extended region and in parts of the southern arms at 3.6\,cm and 6.2\,cm (for the first time), is another significant result of this study. Generally, we found a patchy distribution of the degree of polarization.  The degree of polarization  in the direction of the HII complexes is less than 5$\%$. We also found a weak degree of polarization and some distortion in the regular magnetic field in the central I\,S, which could be linked to the high star forming activity in this region. 

We derived a spectral index map of M33 with a high angular resolution  (90$\arcsec$) \citep[the previous spectral index map was derived at 462$\arcsec$, ][]{Buczilowski_88}. Both the mean total spectral index (0.72\,$\pm$\,0.07, obtained from the 3.6 and 20\,cm maps) and the total spectral index obtained from the integrated spectrum (0.72\,$\pm$\,0.04) are consistent with the result of a statistical study on 56 spiral galaxies by \cite{Gioia}.

The radial profile of the spectral index (Fig.~11) shows not only a general increase but also fluctuations at 2\,$<$\,$R$\,$<$\,4\,kpc, confirming that the decrease in the thermal fraction causes an increase in the spectral index. 
Considering Fig.~13, at 2\,$<$\,$R$\,$<$\,4\,kpc, we found opposite fluctuations in  total intensity and in spectral index. These fluctuations  (in Fig.~13) are  again stronger at wavelengths at which either the thermal emission is stronger (i.e. 3.6\,cm) or the emission correlates well with the thermal emission \citep[i.e. 24\,$\mu$m, see ][]{Tabatabaei_1_07}.  The area at 2\,$<$\,$R$\,$<$\,3\,kpc includes the region between the main arm I (I\,N, I\,S) and the central extended region of the galaxy, corresponding to the small maximum~(bump) in the spectral index profile and a small minimum in the total intensity profile. The area at 3\,$<$\,$R$\,$<$\,4\,kpc includes the arm I\,N and NGC604, corresponding to the small minimum in Fig. 11 and the small maximum in  Fig.~13.


 
Like the star formation radial profile (24\,$\mu$m radial profile, see Sect. 3.6), the mostly non-thermal 20\,cm radial profile is fitted by a larger exponential scale length at $R$\,$<$\,4\,kpc than at $R$\,$>$\,4\,kpc. 
This is a surprising result, indicating that even the non-thermal emission (which is less confined to the star forming regions than the thermal emission) mimics the distribution of its parents, the star forming regions. A spatial coupling between the non-thermal emission and star forming regions was also found in other late-type galaxies \citep{Dumke_00,Chyzy_07}. Therefore, it would be highly interesting to derive the distribution of   the non-thermal emission for a more realistic non-constant spectral index of the non-thermal emission (Tabatabaei et al. paper\,II).

\begin{acknowledgements}
We are grateful to E. M. Berkhuijsen and R. Wielebinski for valuable and stimulating comments. We thank the staff of 100--m Effelsberg telescope and VLA  for their assistance with radio observations.  F. Tabatabaei was supported for this research through a stipend from the International Max Planck Research school (IMPRS) for Radio and Infrared Astronomy at the Universities of Bonn and Cologne. 
\end{acknowledgements}

\bibliography{s.bib}        




\end{document}